# A Century of Science: Globalization of Scientific Collaborations, Citations, and Innovations

Yuxiao Dong
Microsoft Research
Redmond, WA 98052
yuxdong@microsoft.com

Hao Ma
Microsoft Research
Redmond, WA 98052
haoma@microsoft.com

Zhihong Shen
Microsoft Research
Redmond, WA 98052
zhihosh@microsoft.com

Kuansan Wang
Microsoft Research
Redmond, WA 98052
kuansanw@microsoft.com

## ABSTRACT

Progress in science has advanced the development of human society across history, with dramatic revolutions shaped by information theory, genetic cloning, and artificial intelligence, among the many scientific achievements produced in the $20^{th}$ century. However, the way that science advances itself is much less well-understood. In this work, we study the evolution of scientific development over the past century by presenting an anatomy of 89 million digitalized papers published between 1900 and 2015. We find that science has benefited from the shift from individual work to collaborative effort, with over 90% of the world-leading innovations generated by collaborations in this century, nearly four times higher than they were in the 1900s. We discover that rather than the frequent myopic- and self-referencing that was common in the early $20^{th}$ century, modern scientists instead tend to look for literature further back and farther around. Finally, we also observe the globalization of scientific development from 1900 to 2015, including 25-fold and 7-fold increases in international collaborations and citations, respectively, as well as a dramatic decline in the dominant accumulation of citations by the US, the UK, and Germany, from ∼95% to ∼50% over the same period. Our discoveries are meant to serve as a starter for exploring the visionary ways in which science has developed throughout the past century, generating insight into and an impact upon the current scientific innovations and funding policies.

## 1 INTRODUCTION

Science is the discovery, abstraction, and accumulation of knowledge from the universe. It offers the potential to predict, change, and ultimately invent the future. Over the course of human history, the advancement of science has been increasingly responsible for technological and societal evolution. Traced back thousands of years, Aristotle's scientific discoveries and methods profoundly shaped the development of Western civilization. In the modern era, breakthroughs in information theory led to the successive inventions of the telegraph, telephone, television, and Internet, facilitating today's global flow of ideas. Formerly unimaginable innovations in aeronautics and astronautics have further enabled the mobilization of people, lifting us not only across the planet but also, even if only at a nascent stage, into the universe itself.

Notwithstanding the driving force of science in societal development, knowledge concerning the way that science advances itself is sorely lacking. Yet, its advancement is continuously being recorded, with each scientist—past, present, and future—leaving an indelible mark on the process of scientific innovation through his or her research publications, contributing to an ever-expanding body of scientific literature. The digitization of scientific publications over the past century, itself enabled by science, now offers an unprecedented opportunity to study the development of science by using science: the "science of science" [6, 17, 34].

In this work, we study the evolution of science over the past century according to three dimensions. First, we examine the evolving process of collaborations between scientists of different career-ages, institutions, and countries. Second, we characterize referencing behavior over time, with an emphasis on both its individual and collective dynamics. Finally, we investigate the rise and fall of scientific impact across the planet since 1900. Our study is performed on a large-scale scholarly dataset comprised of more than 89 million publications, 795 million citations, and 1.23 billion collaboration relationships spanning from 1900 to 2015, *making this the largest-scale and longest-spanning analysis yet performed on academic data.*

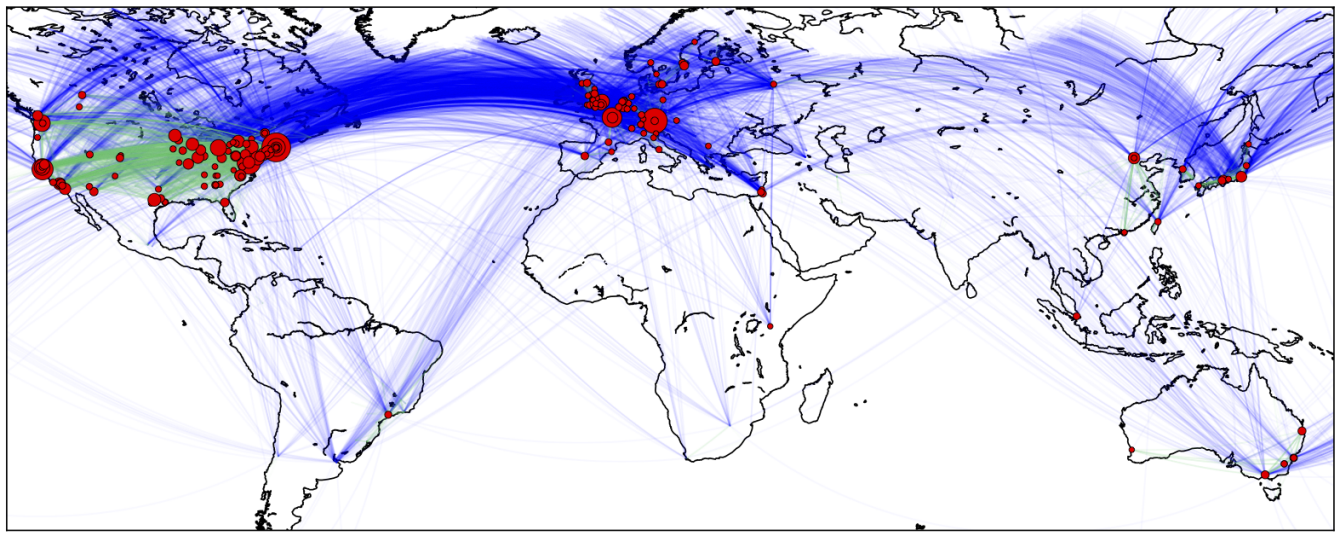

**Figure 1: The planetary-scale view of science between 1900 and 2015.** Blue and green lines represent the *relative* collaboration strength between institutions from different countries and from the same country, respectively. The red circles represent the top 200 most-cited institutions in the world, with Harvard University serving as the top one.

Our study identifies and explores the significant patterns of evolution of the scientific discipline over the $20^{th}$ and $21^{st}$ centuries. Figure 1 shows the planetary-scale view of science, illustrating the effect of the globalization of scientific collaboration and innovations. In general, we find that the volume of scientific publications doubled every 12 years between 1900 and 2015, suggesting an exponential explosion of scientific ideas, despite each scientist's individual productivity remaining surprisingly constant.

- In the context of collaboration, we discover that the size of a publication's author list tripled and the rate of international collaborations increased 25 times over the past 116 years, suggesting an increasingly collaborative scientific production process. We demonstrate that science has benefited from the shift from individual work to collaborative effort, with over 90% of the world-leading innovations (as measured by the top 1% most-cited papers) generated by teams in the 2000s, nearly four times higher than that in the 1900s.
- From the perspective of referencing behavior, we unveil that science has become more open-minded and more broadly shared, as evidenced by a decrease in self-citations among individual authors (30% → 10%) and entire countries (90% → 30%). We also observe that scientists are looking further than ever into the past, citing more and more aged literature, leading to an increasingly visionary scientific discipline.
- Finally, we uncover the diversifying movement of scientific development across the planet over the past century. In the early stage of the $20^{th}$ century, the US, the UK, and Germany were the scientific stalwarts, generating the preponderance of scientific output. The situation of tripartite confrontation was broken down by World War II, leaving the US as the solitary leader in science. Since then, science began a global diversification, with the (re)emergence of scientific advancement achieved by countries from various regions of the world.

Overall, our findings demonstrate that science has increasingly become globally more collaborative, more visionary, and more diverse over time. The patterns we reveal in the evolution of science give rise to important implications for *institutions and governments to better advise and craft research funding policies—such as the encouragement of and support for (international) collaborative projects—and for scientists to discover impactful knowledge, to create revolutionary innovations and, ultimately, to advance the development of science.*

## 2 DATA: MICROSOFT ACADEMIC GRAPH

To comprehensively unveil the evolution of science over the past century, we employ a large-scale scholarly dataset sourced from Microsoft Academic Services [26], i.e., the Microsoft Academic Graph, which has been used in ACM KDD CUP 2016[1] and WSDM CUP 2016[2]. It is composed of more than 100 million publications spanning from 1800 to 2016. From the publications, we construct a collaboration graph with collaboration relationships connecting pairs of authors that have coauthored at least on one paper that was published between 1900 and 2015. Our study is focused on the largest connected component extracted from the original collaboration graph[3]. The resulting data contains 53,326,061 authors and 1,236,796,098 author collaboration links and 795,032,662 paper citation links. Over the course of 116 years, these scientists published 89,559,406 papers. Our overall analysis is conducted on this full data. In addition, for each author, we associate his or her affiliation with geo-location (i.e., latitude and longitude), which determines the corresponding country information. If an institution has multiple locations, its headquarter location is used. For the geographic related analysis, we use 21,695,143 papers (covering 269,690,735

---

[1]https://kddcup2016.azurewebsites.net/
[2]https://wsdmcupchallenge.azurewebsites.net/
[3]The results presented in this work can hold with or without authors who did not have collaborators over their careers.

Table 1: Summary of Questions Answered from the Data.

| Dimension | Questions |
|---|---|
| Collaborations | Q3.1: How many years did it take for science to double its volume? |
| | Q3.2: Did scientists become more collaborative over time? |
| | Q3.3: Did collaborations favor the advancement of science? |
| | Q3.4: How long did scientists maintain their career and collaborations? |
| | Q3.5: How many collaborators and papers did each scientist have? |
| | Q3.6: Where did collaborations happen over the past 116 years? |
| Citations | Q4.1: How many references did scientists cite in a publication? |
| | Q4.2: Had scientists increasingly cited their own papers over time? |
| | Q4.3: How many years back did scientists look in the literature? |
| | Q4.4: Which countries contribute and collect the most citations? |
| | Q4.5: Which countries received more citations than they sent out? |
| | Q4.6: How many times did each country cite themselves vs. others? |
| Impact | Q5.1: How many citations did each paper collect over time? |
| | Q5.2: How many citations did the world-leading research collect? |
| | Q5.3: What are the most influential science institutions in the world? |
| | Q5.4: Which country has the most Nobel laureates for the Nobel Prize? |

paper citations) whose authors are successfully associated with their affiliations' locations.

Given the big scholarly data, our work aims to understand the evolution of science over the past 116 years, with focuses on three dimensions: scientific collaborations, citations, and impact. In specific, along each dimension, we proceed our studies by answering a set of insightful questions, which are summarized in Table 1.

## 3 THE EVOLUTION OF COLLABORATION

Alone we can do so little; together we can do so much.
— Helen Keller, 1920s

In this section, we present an overview of the development of science with an emphasis on the evolution patterns of scientific collaboration, productivity, and research career.

### 3.1 The Growth of Science

• **Q3.1:** *How many years did it take for science to double its volume?* Figure 2(a) provides the yearly number of publications and authors over the past century. The number of scholars who produced scientific publications grew at an exponential rate, doubling every 11 years between 1900 and 2015. As a consequence, the overall volume of publications across all subjects grew exponentially, leading to a twofold increase every 12 years for the past 116 years, which is in accordance with previous discoveries in individual fields such as Computer Science [11] and Physics [25]. The exponential development of science was derailed twice, once around 1915 and once around 1940, corresponding to World Wars I and II, respectively.

### 3.2 Collaboration and Productivity

• **Q3.2:** *Did scientists become more productive and collaborative over the past century?* Despite the rapid expansion in scientific participants and publications, the average yearly number of papers produced per author was fairly constant over the past 116 years at roughly two per year per scientist. However, we also observe that the individual productivity—as measured by the rate between the overall size of publications and authors—dropped by 50 percent from the 1900s to 2000s. The constancy in the yearly average paper count and decrease in individual productivity come from the rising trend of collaborations forged among scientists over time, as evidenced from the increase in the average number of co-authors per paper. Figure 2(c) further details the distribution of papers conditioned on their co-author count over time. Similar to the previously study on a shorter time-frame data [35], we observe that, between the 1900s and 1990s, the share of single-author publications gradually but substantially shrank from 80% to 20%, and in this century, only 15-20% of papers have been authored by individuals. Accordingly, the papers generated by collaboration teams had dominated scientific production since the $3^{rd}$ quarter of the last century. Starting from 1975, in order to solve complex research problems, big collaboration teams with hundreds and thousands of members started to form, which was largely enabled by the development of electronic communications [17]—science itself, such as telegraphy, telephone, and e-mail.

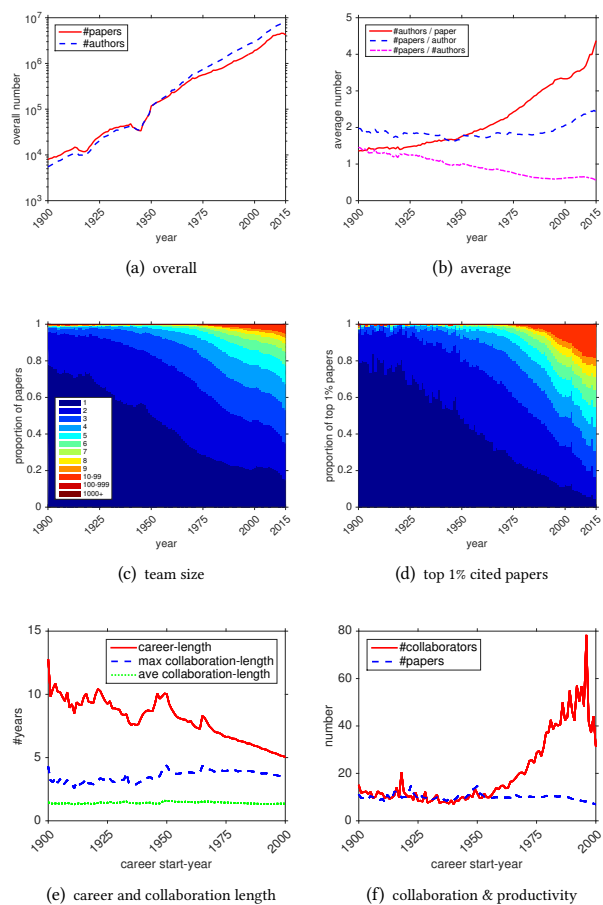

Figure 2: **The evolution of science.** (c) Team size means the number of authors in a paper. (e)(f) The career start-year is defined as the year that one's first paper was published.

• **Q3.3:** *Did collaborations favor the advancement of science?* Indeed, we find that science has benefited from the shift from individual work to collaborative effort. As observed from Figure 2(d), only 20% of world-leading innovations—as measured by the top 1% cited papers—were generated by collaboration teams in the 1900s, while this number increased to over 90% in this century. Further, a random sampled paper would have a 1% probability for being a top 1% cited one, yielding a relative rate $r$=1. We find that in 2015, single-author papers—occupying 15% of the yearly publications—contributed to only 5% of the top papers, suggesting $r$=0.33 (5% / 15%). However, the $r$ for papers with over 10 authors was 4 (20% / 5%)[4], suggesting the positive influence of collaboration on the growth of scientific impact.

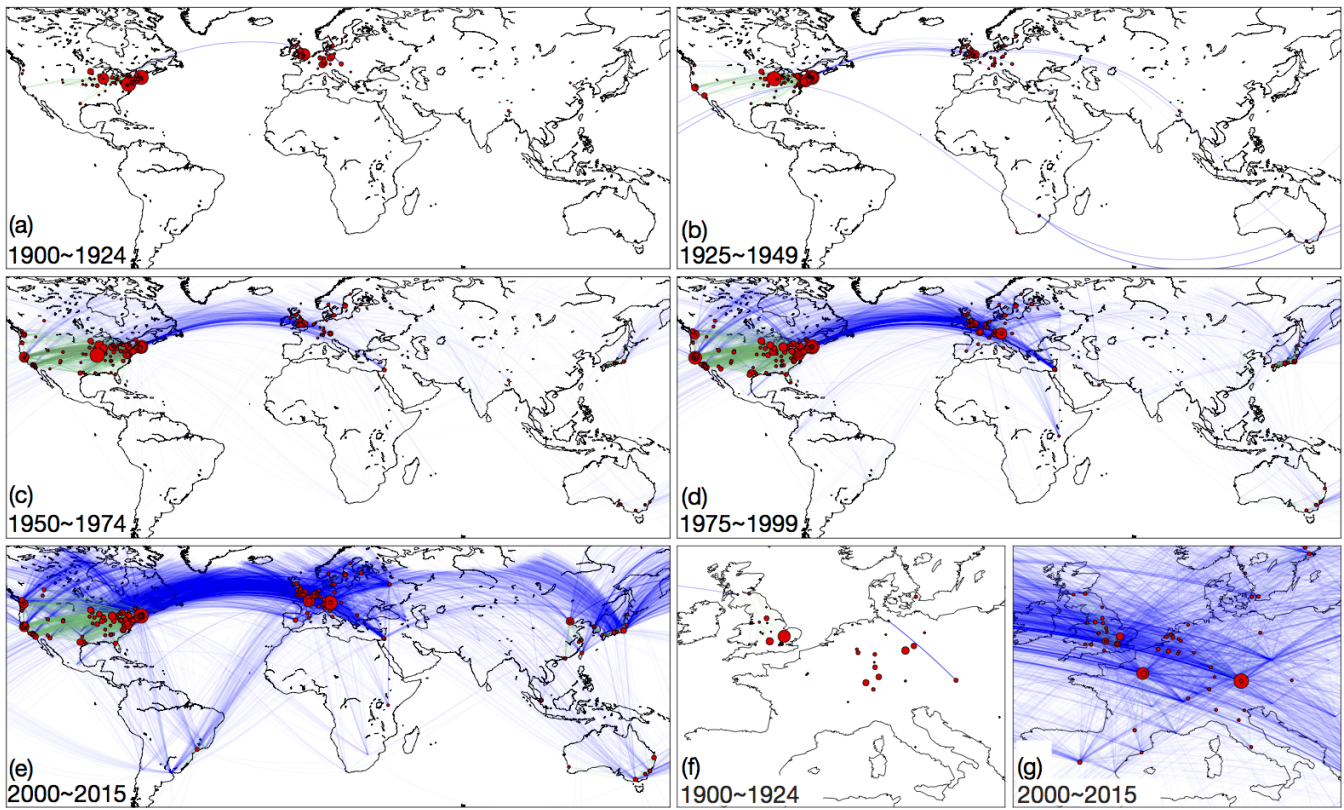

**Figure 3: The landscape of scientific collaborations and impact over the past century.** Blue lines represent the *relative* collaboration strength between institutions from different countries. Green lines represent *relative* collaboration strength between institutions from the same country. The shading density of two types of lines are comparable. The red circles represent the top 200 most-cited institutions in the world, with Harvard University serving as the top one in each period. Each circle's size is proportional to the corresponding institution's citation count.

• **Q3.4:** *How long did scientists maintain their career and collaborations?* To answer this question, we define the start- and end-year of a scientist's career as the dates of the earliest and latest publications attributed to him or her in publication records. Figure 2(e) shows the average of scientists' career length and collaboration duration. Generally, we observe that the length of scientists' careers tended to fluctuate in the first half of the last century on the verge of 10 and then decrease monotonically over the remaining 50 years. We also notice that two major declines in average career length happened to occur during two world war periods. By further breaking down this observation, we find that 40-50% of new scientists never published again after their first-year appearance in publication records. Surprisingly, for all generations of scientists, both the average and maximum length of collaborations has been consistently maintained for 1.5 and 3.5 years, respectively, although their average career length have been declined over the past century.

---
[4]This result can hold with or without self-citations.

• **Q3.5:** *How many collaborators and papers did each scientist have over his or her career?* Figure 2(f) shows the average number of collaborators and co-authored papers each researcher had during his or her career conditioned on the career start-year. Similar to the story revealed in Figure 2(b), we find that the number of collaborated publications during one's career consistently averaged 10, regardless of career start-year, although the number of people he or she collaborated with has been dramatically increased in the second half of the last century. In the first half of the past century, each scientist on average collaborated with 10 people spanning the entirety of his or her research career, while they started to have the tendency to be more and more collaborative over the second half, ending up with more than 60 collaborators if their careers started in in the 1980s. Notice that the decreasing collaborator number in the late 1990s results from the fact that scientists who started to publish at that time have not yet ended their careers as of 2015—the last year represented in our dataset.

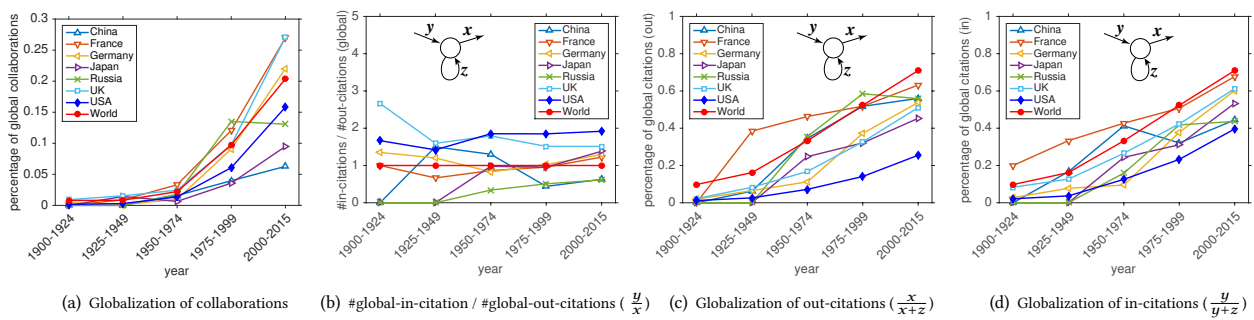

Figure 4: **Globalization of collaborations and citations.** (b)(c)(d) Take a country as a research unit, $x$ denotes the number of citations she cites others, $y$ denotes the number of citations she collects from others, and $z$ denotes the number of citations she sends to (receives from) herself.

## 3.3 Collaborations Across the World

● **Q3.6:** *Where did collaborations happen over the past 116 years?*
The observations above demonstrate that science has developed itself into a much more collaborative enterprise over the past century. Figure 3 plots the world maps embedded with two types of collaborations, including those across institutions within the same country (green lines) and those across institutions of different countries (blue lines).

Between 1900–1924, collaborations across different institutions predominantly occurred among American institutions; those across different countries were dominated by the connections between the US and the UK. Both types of collaborations were relatively weak. The most inner-collaborative institution was Harvard University during this period. Between 1925–1949, in addition to the bond between the US and the UK, international collaborations started to form between India and the UK, as well as between Australia and the US. Due to World War II, collaborations in Europe overwhelmingly shrank during this quarter. Meanwhile, in addition to the Northeastern region, collaborations in America were rapidly developed in the Midwest (the Great Lakes Region). Between 1950–1974, Israel and Japan started to participate in international collaborations. At the same time, inside the US, the west coast and the south region became strong areas of scientific collaboration. Between 1975–1999, Africa began to develop collaborations with Europe. Surprisingly, the inner-institution collaborations within America decreased relative to those in Europe, although the absolute number of collaborations grew substantially over time for all countries. In this century, more and more countries have risen to the international stage to collaborate extensively with others.

Additionally, Figure 4(a) reports the evolution of global collaborations by considering each country as a research unit. We find that during the first quarter of the $20^{th}$ century, more than 99% of the collaborations were forged within the same country, as indicated by the red line. That is, fewer than 1% of all scientific collaborations happened across geographic boundaries. However, the frequency of global collaborations has been consistently growing over time. By the end of the last century, the share of international collaborations had reached about 10%. As of this century, the dominance of intranational collaborations has further dropped to 80%, with 20% of publications coauthored by scientists from different countries.

We notice that as of the $21^{st}$ century, the three EU developed countries—the UK, Germany, and France—have higher international collaboration rates than the global average, while the USA, Russia, Japan, and China have lower rates. Interestingly, Russia's current international collaborations are weaker than those during the last quarter of the $20^{th}$ century. Our detailed analysis suggests that the strong international collaborations exhibited by the three EU countries lie in the various bonds forged between EU countries as well as those between EU and the US. Similarly, US institutions have very strong ties with each other, while institutions in Japan tend to collaborate internally.

## 3.4 Summary

From the perspective of collaboration, we find that the average length of a publication's author list **tripled** between 1900 and 2015, suggesting an increasingly collaborative scientific process. We discover that the rate of international collaborations has increased **25 fold** over the past 116 years, revealing the globalization of scientific collaborations. We also find that in the 2000s, over 90% of the world-leading innovations were generated through collaborations, nearly **four times** higher than that in the 1900s, demonstrating that the scientific discipline has benefited from a growth in collaborative effort that is increasing needed for producing technological innovations in the modern world.

## 4 THE EVOLUTION OF CITATION

*If I have seen further, it is by standing on the shoulders of giants.*
— Isaac Newton, 1676

Science is largely inspired by and developed upon previously discovered knowledge. This cumulative nature of scientific production is represented in the form of scientific citation. Citing literature not only acknowledges others' credit but also establishes the new work's authority that is collectively borrowed from prior work. Here, we study the evolution of citing behavior in science over the past 116 years. Specifically, we seek to achieve a clear understanding of citing dynamics, both individually and collectively.

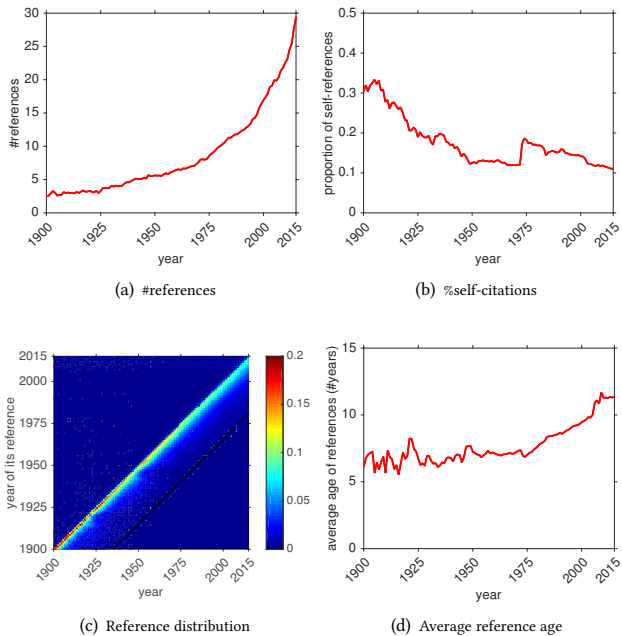

(a) #references  (b) %self-citations

(c) Reference distribution  (d) Average reference age

**Figure 5: The evolution of referencing behavior.** The age of a reference is defined as the time between its date of publication and the date of publication of a paper that cites it. A citation is a self-citation if the cited publication shares at least one common author with the publication that cites it. (c). The sum of each column is equal to 1.

## 4.1 Reference Characterization

We characterize the dynamics of individual reference behavior by answering the three questions below.

• **Q4.1:** *How many references did scientists cite in a publication?* Figure 5(a) shows how the average size of a paper's reference list changes from 1900 to 2015. Similar to the previous discovery in Physics [25], in the 1900s and 1910s, each scientific paper on average had only two to three references. However, the size of a reference list grew in a superlinear fashion from the 1900s to the present day, which is different from the linear increase in Physics. As of 2015, modern scientists reference on average 30 papers in a publication. The drastic increase in the typical number of references is in part due to the accumulation of available literature that has grown exponentially; it may also partially result from the much more complex, interdisciplinary nature of modern science that has developed over time.

• **Q4.2:** *What if scientists had increasingly cited their own papers over time?* This could have been also responsible for the boost of a paper's reference list size. Figure 5(b) provides the average proportion of self-citations in a publication. It turns out that scientists' tendency to cite their own papers has fallen over time, as evidenced from a self-citation rate of over 30% in the 1900s and only 10-15% in the 2000s. In the 1950s, the rate of self-citations had actually dropped to only around 10%, and was stable until 1975—the year that Thomson Reuters started to release the Journal Citation Reports (JCR) [4]—from which we observe an immediate jump in the

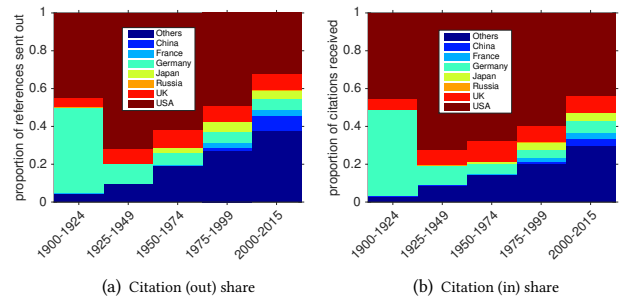

(a) Citation (out) share  (b) Citation (in) share

**Figure 6: The evolution of reference and citation shares.** (a) The share of citation relationships that come from different countries; (b) The share of citations that are collected by different countries.

self-citation rate from 10% to 20%, possibly suggested by journal organizers to improve rankings.

• **Q4.3:** *How many years back did scientists look in the literature?* Figure 5(c) shows the references' age distribution of papers published in different years. The highlighted diagonal line indicates that across all generations in the last century, most inspiration came from publications published within five years, suggestive of the iterative nature of science. Further, the bright diagonal stripe (bounded by the black line) suggests that scientists can be inspired by the work published as far back as 25 years ago, explaining the gradual fade of scientific novelty [34] and the transition of novel discoveries into common knowledge—and into not being cited—over time.

We also find that color of the diagonal line changes from red at the bottom-left corner to cyan at the top-right, suggesting that science has, over the past 116 years, evolved from myopic referencing (i.e., referencing "young" papers) to deep referencing (i.e., referencing "old" papers), which is also evidenced from the gradual increase in the average reference age of papers shown in Figure 5(d). The average age of a paper's references stayed at around 6 years until 1975, after which it began to slowly increase over the next 40 years, finally reaching 11 years in the 2010s. We posit that the gradual transition to deep referencing between 1975 and 2015 was the synergistic effect of 1) the start of peer review in the 1960s (suggested for the field of Physics by [25]); 2) the release of various venue rankings starting from 1975, resulting in a more accurate distribution of credit to older literature; and 3) the development of digital publication databases, providing an easier way to track and trace more publications—especially the old ones.

## 4.2 Citing Across Countries

We examine the flow of references by using the global average and seven representative countries mentioned above. The visualization of references at the planetary scale is skipped due to page limit.

• **Q4.4:** *Which countries contribute and collect the most citations?* Figure 6 shows the distribution of citations that were sent out to and collected from the world. During the $1^{st}$ quarter of the $20^{th}$ century, the US, together with Germany, created the dominant references and collected the dominant citations. The balance between Germany and the US was broken in the $2^{nd}$ quarter, largely due to the influence of World War II, propelling the US to a powerful lead in scientific development throughout the world. From then on, the UK and Germany have kept relatively consistent shares of scientific participation to the current day. Since the second half of the century, Japan and France, together with the remaining hundreds of countries, started to play considerable roles in scientific production, with the US's consistent loss of both scientific import and export shares. Starting from the $4^{th}$ quarter, China marched to contribute to science extensively. However, compared to the relatively large reference shares that China and the remaining 100+ countries contributed to, the shares of citations they collected were noticeably low. In contrast, the share of citations the US attracted from others was larger than share of references sent out to others. In summary, during the early $20^{th}$ century, the US, Germany, and the UK created 95% and collected 97% of the world's citations, while these two shares were decreased by about half as of the $21^{st}$ century, to 46% and 58%, respectively. Altogether, throughout the $20^{th}$ and $21^{st}$ centuries, science has indeed developed into an open world, facilitating associations between increasingly diverse participants and accommodating a steadily shrinking polarization.

• **Q4.5:** *Which countries received more citations than they sent out?* Figure 4(b) shows the rate between the number of citations one country receives and the number she sends out, wherein it is equal to 1 for the world as a whole (red line). Overall, we can see that over the five periods, the US and the UK were consistently above the equal (red) line, while Russia was invariably below the line, with both France and Germany close to the line. From the perspective of evolution, however, Russia's rate increased over time, in contrast to a decreasing trend of the UK. The rate of the US had been around 2 for the past 116 years, which means that the number of citations the US collected from the world was as twice as many she cited others, indicating that the US was the universally attractive science producer relative to the other six.

• **Q4.6:** *How many times did each country cite themselves vs. other countries?* Figures 4(c) and 4(d) show the percentage of total citation volume that each representative country contributes to and collects from other countries. We find that during the first quarter of the $20^{th}$ century, all seven representative countries sent out fewer than 2% of their citations to others, while fewer than 10% of their citations were received from others (excluding France). This indicates a strong referencing homophily effect at the country level (>98%), partially due to the relative inconvenience of scientific communication at the time. The dominance of referencing homophily has dramatically weakened since the second half of the $20^{th}$ century, although the US, Germany, the UK, and Japan still have a relatively strong preference to citing research within their own country. As of this century, all seven countries—and the US in particular—send and collect fewer than 70% (global average) of their citations to and from the broader world, suggesting they have a stronger tendency than other countries for intranational citations. Nevertheless, we have witnessed the development of a substantially more diverse citation world over the past 116 years. Overall, we find that only 10% of citations were sent out to scientists from different countries in the first quarter of the $20^{th}$ century, while this share has dramatically increased to 70% in this century.

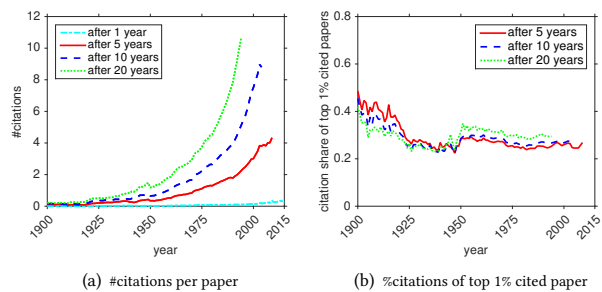

(a) #citations per paper  (b) %citations of top 1% cited paper

**Figure 7: The evolution of citation counts.**

## 4.3 Summary

Concerning citation behavior, we unveil that scientists are looking further than ever into the past, citing more and more aged literature, as evidenced by a **two-time** increase in the average age of a paper's references from 1900 to 2015, leading to an increasingly visionary scientific discipline. We demonstrate that science has also become more open-minded and more broadly shared, as evidenced by a decrease **from 30% to 10%** in self-citations among individual authors and **from 90% to 30%** among entire countries.

## 5 THE EVOLUTION OF IMPACT

Science knows no country, because knowledge belongs to humanity, and is the torch which illuminates the world.

— Louis Pasteur, 1876

In this section, we examine the evolution of scientific impact over the past 116 years. We present the macro trends of scientific impact dynamics, as well as an emphasis on geographic and country-level analysis.

## 5.1 Inflation, Dominance of Top Research

• **Q5.1:** *How many citations did each paper collect over time?* Figure 7(a) shows the average citation counts received by each paper within the time-frame of 1, 5, 10, and 20 years. We observe that the number of citations each paper collected within 5 to 20 years increased dramatically over the past century. Papers published before 1950 collected on average less than two citations after 20 years, while this number increased to four citation for papers published before 1975 and further reached ten for those published before 1995. This could be explained as the consequence of the explosive growth in the size of reference list (Figure 5(a)), as well as the gradually deepening referencing behavior. The inflation of citation counts over time gives rise to critical implications for the evaluation of publications generated at different time periods and as a result, scientists at different career-ages.

• **Q5.2:** *How many shares of citations did the world-leading research collect over time?* Figure 7(b) presents the citation shares accrued by the top 1% most-cited publications each year. We observe that, overall, ~25% of the citations that were received by papers published in the same year went to the top 1%. The sharpest polarization happened in 1900s (40–50%). A previous case study [7] has also shown a similar trend, wherein the top 1% most-cited papers in five journals published in 1990 collected 5%~20% of the citations within 5–20 years. In addition, we find that before the 1940s, the top 1% most-cited papers collected a larger share of citations within a short time-frame (e.g., 5 years) than a long-term period (e.g., 20 years). Since the middle of the $20^{th}$ century, this situation has been reversed, with top research likely to generate more impact in a longer time-frame than in a shorter one. This reversal is a result of the deep referencing trend revealed in Figures 5(c) and 5(d).

Table 2: The rate between #citations of the top 1 institution and #citations of the $200^{th}$ one over the past 116 years.

| year | 1900-1924 | 1925-1949 | 1950-1974 | 1975-1999 | 2000-2015 |
|------|-----------|-----------|-----------|-----------|-----------|
| rate | 302 | 80 | 25 | 15 | 11 |

## 5.2 Distribution of Top Institutions

Based on our dataset, as of 2015, about 19,000 institutions have published scientific papers, including universities, research institutes, and industry labs. A natural and interesting question arises:

• **Q5.3:** *What are the most influential science institutions in the world?* Figure 3 visualizes the top 200 most-cited institutions, as measured by citation counts across five time-frames. The size of a circle is proportional to the citation counts received by the corresponding institution. Institutions ranked beyond 200 are hardly observable from the map and hence skipped plotting.

Between 1900–1924, the top four world innovation centers—as measured by citation counts—were Harvard U., Johns Hopkins U., U. of Cambridge, and U. of Chicago. Undoubtedly, the US, the UK, and Germany led the world in science, as most of top-cited institutions in the first quarter of the $20^{th}$ century were located in their territories.

Between 1925–1949, the US institutions continued dominating citation collections, with high margin, not only having all top four research institutions—Harvard, U. of Chicago, Columbia U., and Massachusetts Institute of Technology (MIT), but also incubating the emergence of top institutions beyond the northeastern region, such as Stanford U. and California Institute of Technology on the west coast. In contrast to the dominance of scientific impact by the US, top institutions in Continental Europe drastically shrank both quantitatively (#red-circles) and qualitatively (the size of red-circles), due to the catastrophic damage of World War II. In addition, the War had a significantly negative impact on the research development in the UK as well.

Between 1950–1974, the US was still the dominant force in science, followed by the UK. From this quarter, top institutions started to emerge in North Europe, Israel, Japan, and Australia. It is worth noting that the world top seven cited institutions—those top places were all relatively close—were Harvard, Washington University in St. Louis, Bell Labs, MIT, UC Berkeley, National Institutes of Health (NIH), and Stanford. This was the harbinger that industry labs and government institutes had became a strong force in the scientific enterprise.

Between 1975–1999, Continental Europe had fully recovered from World War II. Germany's Max Plank Society (MPS) was ranked as the $5^{th}$ most-cited institution in the world during this quarter, right after Harvard, Stanford, NIH, and MIT.

As we step into this century, MPS furthers its ranking to the $2^{nd}$ place and French National Center for Scientific Research (CNRS) moves to the $4^{th}$ place, making Europe again as competitive as the US in science, with Harvard as the $1^{st}$, NIH as the $3^{rd}$, and Stanford as the $5^{th}$. More importantly, the map also demonstrates that the distribution of the top 200 research institutions has been significantly diversified in this century: China, South Korea, and Singapore in Asia, Brazil in South America, and many more places in Europe are joining in this top list. In addition, the gap between institutions in Asia and Australia and the top-most institution in the world—Harvard—has decreased from the last century to the current, demonstrating the rapid rise of scientific impact in the Asia-Pacific region during the past 16 years.

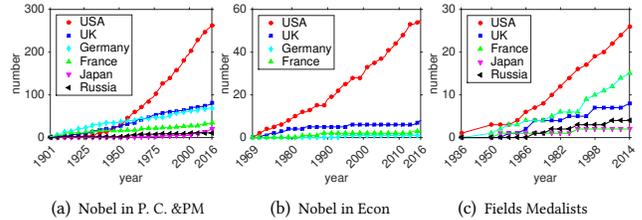

(a) Nobel in P. C. &PM    (b) Nobel in Econ    (c) Fields Medalists

Figure 8: The distribution of major scientific award winners. (a) The number of Nobel Prize Laureates in Physics, Chemistry, and Physiology or Medicine by Country; (b) The number of The Sveriges Riksbank Prize in Economic Sciences in Memory of Alfred Nobel by Country; (c) The number of Fields Medalists by Country. As to the **A.M. Turing Award**, 50 of all 64 winners between 1966 and 2015 come from the US.

Further, Table 2 shows that the gap between the citation counts collected by the top 1 institution and the top $200^{th}$ institution has consistently and drastically decreased over the past five periods of time, another illustration of a more diverse development in science over time.

Additionally, from Figures 3 (f) and (g), we find that in the early stage of the $20^{th}$ century, U. of Cambridge and U. of Oxford in the UK were undoubtedly the top two most influential research institutes in Europe. As of the $21^{st}$ century, both are still relatively competitive to most other places in Continental Europe, but not to MPS and CNRS anymore. The two in Germany and France have significantly surpassed them as measured by citation counts, although both countries suffered from much worse devastation to science than the UK during World War II.

Overall, we witness the global diversification of scientific innovations from the absolute dominance of the Northeastern US, the UK, and Germany in the 1900s to the shared leadership of both US coasts and Continental Europe in the second half of $20^{th}$ century to the rapid rise of research in Asia and the other continents in the $21^{st}$ century.

## 5.3 Distinguished Award Winners

In addition to using citation counts to measure impact, we further investigate the country distributions of recipients of distinguished scientific awards[5] over time, including the Nobel Prize Laureates in Physics, Chemistry, and Physiology or Medicine [5], The Sveriges Riksbank Prize in Economic Sciences in Memory of Alfred Nobel [1], the Fields Medal (Math) [3], and the A.M. Turing Award (Computing) [2]. Simply, we try to answer the following question:

---
[5] The recipient's country is determined from the official award website.

- **Q5.4:** *Which country has the most laureates for the Nobel Prize, Fields Medal, and Turing Award over the awards' histories?* Figure 8 shows the cumulative number of award winners to date since the inception of these awards. From each figure, we observe the dominance of the US since the 1960s in each of these distinguished awards. Before the 1950s, however, Germany had the most number of Nobel Laureates in Physics, Chemistry, and Physiology or Medicine, and both the UK and France had more winners than the US (Figure 8(a)). After World War II, the US started to dominate the Nobel Prize in these three disciplines (263 winners as of 2016), leaving a large gap between it and the next countries—the UK (82) and Germany (69). As for the Nobel Prize in Economics, the Fields Medal, and the Turing Award, the US has dominated them since their inception. As of 2016, 55 Nobel laureates in Economics and 26 Fields Medalists come from the US, followed by 8 from the UK and 15 from France. Between 1966 and 2015, the A.M. Turing Award had been awarded to 64 computer scientists, with 50 of them from the US, four from the UK, two from Israel, two from Norway, and the remaining six from six different countries. We also find that after World War II, Japanese scientists started to make scientific breakthroughs in Physics, Chemistry, Physiology or Medicine, and Mathematics, as testified by the considerable number of award recipients from Japan since then.

## 5.4 Summary

From the perspective of impact and innovations, we witness the global diversification of scientific development across the planet over the past century. In the early stage of the $20^{th}$ century, only **4%** of world-leading (as measured by top 200 most-cited) research institutions were located outside the US, the UK, and Germany. The situation of tripartite confrontation was broken down by World War II, leaving the US as the solitary leader in science for half a century. Since then, science has begun a gradual global diversification, now with **40%** of the innovation centers located in countries dispersed across various regions of the globe.

## 6 RELATED WORK

Science of science has been an emerging discipline wherein science is used to study the development of itself—science [6, 12, 24, 31]. Traditionally, a significant body of work has been focused on designing fair and effective scientific metrics to quantify the impact of publications, individuals, venues, and institutions, such as impact factor [12], *h*-index [15], and others [9, 16, 28]. Given a practical measurement of scientific impact, scientists started to look at the potential to predict the future of science. In particular, the prediction of citation count and *h*-index has attracted various of attention from diverse communities, including information science [6], computer science [13, 19, 36], Physics [34] and so on. In addition to the prediction of impact, tremendous effort has been devoted to the characterization of scientific impact, such as its universal citation distributions [22], author credit allocation [23, 24], and *h*-index's growth pattern [10]. Further, network science has brought a network perspective to the analysis of scientific collaboration and citation, including the anatomy of collaboration [20, 29] and citation [21] networks, team organization [17], and heterogeneous entity and structure [27, 33]. Besides the study of science, noticeable academic and scientific online systems, such as AMiner [30], CiteSeerX [14], Google Scholar [32], and Microsoft Academic Services [18, 26], have been developed for better servicing the scientists for better science.

Herein we aim to unveil the evolution of science over a long timeframe, that is, 1900-2015, leading to a crucial distinction between this work and most of literature wherein the focus is on either the static analysis or the dynamic view within a couple of years. A very recent work examined the progress of Physics research over the $20^{th}$ century [25]. The major differences between ours and this work lie in that 1) in addition to analyze the growth, we propose to focus on a new dimension—the influence of geography and country across the planet on science evolution, and 2) we analyze a much bigger scholarly data covering all fields of science. Another two related work studied the collaboration or citation links happened between two institutions conditioned on their geographical locations by using a set of papers published between 2003 and 2010 [21], and between 1996 and 2013 [8], respectively, disabling the potential to unveil the evolution of science over centuries. Our work further differs from it by examining science from multiple broad views rather than their interest in geo-distance and science.

## 7 CONCLUSION

In this work, we present an anatomy of science spanning the $20^{th}$ century and the first 15 years of the $21^{st}$ century. Our study provides evolutionary and planetary-scale views of scientific development. In the $1^{st}$ quarter of the last century, scientific development was led by the US, the UK, and Germany. In this period, science tended to rely more on individual scientists' efforts and talents. The Second World War left science in Europe overwhelmingly damaged leading into the $2^{nd}$ quarter of the century, including in both Germany and the UK. From this period, science began a gradual, continual process of increasing collaboration, openness, and diversity. The average number of collaborators per scientist has at least quintupled; the age of the literature that scientists look back on and cite has increased by 90-100% and transitioned away from a focus on their own work to the work of others; and the countries that advance science have undergone a radical geographical diversification. Our findings unveil the evolutionary patterns of science over 116 years. This knowledge—the knowledge of science itself—has the potential to help scientists improve the scientific discipline and, as a result, to forge a better world.

Despite our extensive and manifold analysis of big scholarly data, there are still several remaining dimensions for characterizing science in the future. First, while this work focuses on all fields of science in an aggregated way, it would be interesting to break down science into different disciplines and topics, enabling us to answer questions like: Which countries have been the leaders in deep learning, big data, or artificial intelligence over time? Second, publication venues, such as journals and conferences, are the medium of science, making it necessary to track the evolution of venues and their impact on scientific development. Finally, it would be natural to investigate the interplay between economic development and science advances.

**Acknowledgments.** We would like to thank Reid Johnson for suggestions and comments.